\documentclass[twocolumn,journal]{IEEEtran}
\usepackage[T1]{fontenc}
\usepackage[latin9]{inputenc}
\usepackage{verbatim}
\usepackage{float}
\usepackage{units}
\usepackage{textcomp}
\usepackage{amstext}
\usepackage{amssymb}
\usepackage{graphicx}
\usepackage[unicode=true,
 bookmarks=true,bookmarksnumbered=true,bookmarksopen=true,bookmarksopenlevel=1,
 breaklinks=false,pdfborder={0 0 0},pdfborderstyle={},backref=false,colorlinks=false]
 {hyperref}
\hypersetup{pdftitle={Your Title},
 pdfauthor={Your Name},
 pdfpagelayout=OneColumn, pdfnewwindow=true, pdfstartview=XYZ, plainpages=false}

\makeatletter

\newcommand{\lyxmathsym}[1]{\ifmmode\begingroup\def\b@ld{bold}
  \text{\ifx\math@version\b@ld\bfseries\fi#1}\endgroup\else#1\fi}

\providecommand{\tabularnewline}{\\}

\let\oldforeign@language\foreign@language
\DeclareRobustCommand{\foreign@language}[1]{%
  \lowercase{\oldforeign@language{#1}}}

\usepackage[caption=false,font=footnotesize]{subfig}

\makeatother

\begin{document}
\title{Theoretical Limits of Backscatter Communications}
\author{Clemens~Korn\thanks{Clemens~Korn is with the Dependable M2M Research Group, Technische
Universitaet Ilmenau, Germany, and with the Fraunhofer IIS, Fraunhofer
Institute for Integrated Circuits IIS, Nuremberg, Germany, e-mail:
\protect\href{mailto:clemens.korn@tu-ilmenau.de}{clemens.korn@tu-ilmenau.de}},~Joerg~Robert\thanks{Joerg~Robert is with the Dependable M2M Research Group, Technische
Universitaet Ilmenau, Germany, and with the Fraunhofer IIS, Fraunhofer
Institute for Integrated Circuits IIS, Erlangen, Germany, e-mail:
\protect\href{mailto:joerg.robert@ieee.org}{joerg.robert@ieee.org}},~Tobias Dräger\thanks{Tobias~Dräger is with the Fraunhofer IIS, Fraunhofer Institute for
Integrated Circuits IIS, Self-Powered Radio Systems Department, Erlangen,
Germany, e-mail: \protect\href{mailto:tobias.draeger@iis.fraunhofer.de}{tobias.draeger@iis.fraunhofer.de}}}
\IEEEaftertitletext{This work has been submitted to the IEEE for possible publication.
Copyright may be transferred without notice, after which this version
may no longer be accessible.}

\maketitle
\begin{abstract}
Backscatter communication is a hot candidate for future IoT systems.
It offers the possibility for connectivity with tiny amounts of energy
that can be easily obtained from energy harvesting. This is possible
as backscatter devices do not actively transmit electromagnetic waves.
Instead they only reflect existing electromagnetic waves by changing
the antenna load. This fact leads to significant differences compared
to classical communication wrt. the modulation schemes and achievable
data rates. However, to our best knowledge nobody has so far systematically
analyzed the achievable data rates and transmit ranges for different
parameter configurations. Within this paper we derive theoretical
bounds for backscatter communications based on classical information
theory. We then use these bounds to analyze how different parameters
\textendash{} e.g. the distance, the frequency, or the transmit power
\textendash{} affect the achievable data rates. The bounds are derived
for mono-static configuration, as well as for bi-static configurations.
This allows feasibility analyses for different use-cases that are
currently discussed in 3GPP and IEEE 802.
\end{abstract}

\begin{IEEEkeywords}
Backscatter communication, IoT, low-power communication, ambient backscatter,
wireless sensor networks
\end{IEEEkeywords}

\IEEEpeerreviewmaketitle{ }

\section{Introduction}

\renewcommand{\baselinestretch}{0.9657}\normalsize

Backscatter-based radio communication is already a proven and well-established
technology, e.g. in UHF-RFID (Ultra High Frequency Radio Identification)
\cite{BackscatteringTorres2021}. However, it is also a hot candidate
for future IoT (Internet of Things) systems \cite{xu2018practical}.
This is particularly the case for passive IoT systems and sensors.
These systems are typically not equipped with a powerful battery and
have to harvest their complete energy. Hence, they can only afford
to spend very tiny amounts of energy for RF (Radio Frequency) communication.
These passive devices are currently discussed for future 6G communication
networks \cite{9145564,Nawaz2021}, and as a topic for a possible
new 3GPP Study Item ``Ambient IoT''. Also in IEEE 802.11 (WiFi)
there currently exists the ``Ambient Power (AMP)'' Technical Interest
Group.

Backscatter devices are not an active transmitter of electromagnetic
waves. Instead, they are reflecting already existing electromagnetic
waves by changing the impedance of their antenna, e.g. they switch
the antenna load between open and short \cite{FinkenzellerEnglisch}.
This leads to a very tiny power consumption for the backscatter devices,
mainly due to two reasons: First, no amplifier is required inside
the backscatter device. Secondly, the backscatter device does not
need any high-precision oscillator. It does not have to generate a
high frequency signal at the carrier frequency, but only a modulation
at the symbol rate, which normally is several magnitudes smaller than
the carrier frequency. An example for the power consumption of a backscatter
modulator is e.g. presented in \cite{BackscatterModulatorEnergieverbrauch}.
The power consumption only depends on the data rate, but not on the
carrier frequency. This shows, that backscattering is particularly
suited for all kinds of use-cases where small amounts of data have
to be transmitted over a limited range and where low power consumption
and low complexity at the transmitter side is required.

The signal that is backscattered and modulated by the passive device
can either origin from a dedicated source, or from an ambient source.
Relying on a dedicated source is the traditional approach for backscattering.
Here, a signal (typically a continuous wave) is radiated with the
aim of providing a carrier signal that can be backscattered by a device.
Usually, the signal from the dedicated source is also used for powering
the backscatter device, as done in case of passive UHF-RFID \cite{Spec}.
Another approach that recently got increasing attention in research
is to use an ambient source \cite{AmbientBackscatterSurveyHuynh}.
Here, the backscatter device uses already existing signals from other
radio systems that have not been radiated for backscattering purposes.
Examples are radio broadcast or signals from mobile cellular networks.

There already exists a plethora of work where ambient backscattering
has been implemented. In \cite{8057162}, measurement results are
presented for ambient backscattering using signals from FM Radio,
ATSC Digital TV, CMDA2000, or UMTS. In \cite{Talla2017}, LoRa signals
were used while \cite{Chen2021} uses Bluetooth signals. FM broadcast
signals were e.g. used in \cite{201552,8058860,9295547}, while WiFi
signals were used in \cite{9513239,KelloggWIFIBackscattering,9488716}.
Finally, generic OFDM (Orthogonal Frequency Division Multiplex) signals
were used in \cite{8647245,8103807}.

Further, there are papers that theoretically investigate some aspects
of the expected performance of backscattering. The link budget is
analyzed in \cite{LinkBudgetPark2021} and \cite{8885700}, while
\cite{7948789} provides some insight on the achievable data rate
as a function of the signal-to-noise ratio (SNR). Considering multiple
users, the channel capacity and outage performance is investigated
in \cite{AmbientBackscatterZhao2018,9051982} and \cite{9055221}.
However, to our best knowledge no detailed extensive theoretical analysis
on the effects of different parameters has been performed so far.

Therefore, in our work we investigate the theoretical limits of backscattering
communications in terms of maximum achievable payload data rate based
on Shannon's communication theory. This paper is structured as follows.
Sec. \ref{sec:Channel-Model} presents the channel model with two
different backscattering configurations, a bi-static and a mono-static
one. Based on these, the theoretical bounds of backscatter based communications
are derived in Sec. \ref{sec:Derivation-of-the-Theoretical_Bound}.
In Sec. \ref{sec:Influence-of-Various}, we show how various system
parameters influence the bound for the data rate. Finally, Sec. \ref{sec:Application-of-the-RIFD}
shows a practical example and Sec. \ref{sec:Conclusion} concludes
this document. 

\section{Backscatter Channel Model\label{sec:Channel-Model}}

In this section, we present our channel model for the later following
investigations. This includes the backscattering setup, the path loss
model, and the noise.

\subsection{General Assumptions}

We are focusing on the theoretical bounds. Consequently, we always
assume perfect conditions, which may not be reachable in real systems.
We also do no care about the type of signal that we use for the backscatter
modulation. It is only defined by its power $P_{C}$. Further, we
assume a correlation receiver \cite{Proakis2007}. Throughout this
document the passive device backscatters $\unit[100]{\%}$ of the
received signal power, which corresponds to a backscatter efficiency
of $\mu=1$. This also means that we ignore potential losses caused
by the energy harvesting of the device, as e.g. done by passive UHF-RFID
systems \cite{FinkenzellerEnglisch,Kuester2013HowGood}. Further,
we assume that the signals are transmitted via line-of-sight and that
RX and TX antennas are always perfectly directed towards each other
during the communication. Finally, we assume that the performance
of the receiver is not affected by the original carrier signal generated
by the source, which is a typical problem in practical scenarios \cite{FinkenzellerEnglisch}.

\subsection{Backscatter Configurations}

Generally, there exist two different backscattering configurations,
i.e. the bi-static and the mono-static configuration.

\subsubsection{Bi-static Configuration}

\begin{figure}[t]
\includegraphics[width=3.5in]{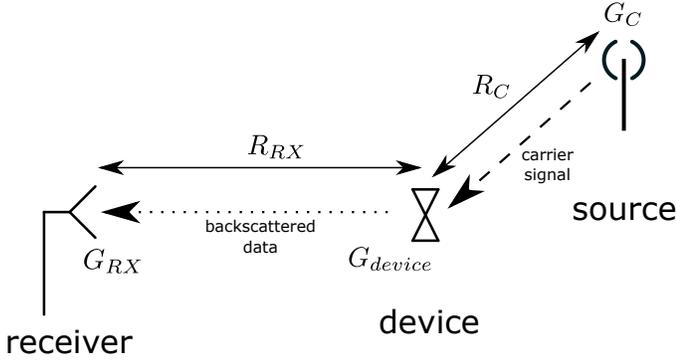}

\caption{Channel model of the bi-static configuration: The path of the carrier
signal from carrier signal source to backscatter device is illustrated
by the dashed arrow, while the path of the backscattered signal from
backscatter device to the receiver is illustrated by the dotted arrow. }
\label{fig:channel_model_bistatic}
\end{figure}
Fig. \ref{fig:channel_model_bistatic} shows the bi-static configuration.
A high-frequency radiator (source) transmits the carrier signal to
the backscatter device. The transmit power of the carrier source is
given by $P_{C}$. The backscatter device then backscatters the signal
to the receiver. The configuration is called bi-static, as the source
and receiver do not use the same antenna.

The bi-static configuration is the typical configuration for ambient
sources, as the source could be a cellular network station. However,
it can also be used in case of dedicated sources for the backscattering
signal, as it does not suffer due to problem of simultaneous transmission
and reception from the same antenna. 

In the following, $R_{C}$ denotes the distance from the source to
the backscatter device, and $R_{RX}$ denotes the distance between
the backscatter device and the receiver. Further, the antenna gain
of the source is given by $G_{C},$ $G_{device}$ defines the antenna
gain of the backscatter device, and $G_{RX}$ is the antenna gain
of the receiver.

A relevant factor for the further calculations is the EIRP (effective
isotropic radiated power) $P_{C,EIRP}=P_{C}G_{C}$ of the source,
as the maximum EIRP is typically limited by the frequency regulation.

\subsubsection{Mono-static Configuration}

\begin{figure}[t]
\includegraphics[width=3.5in]{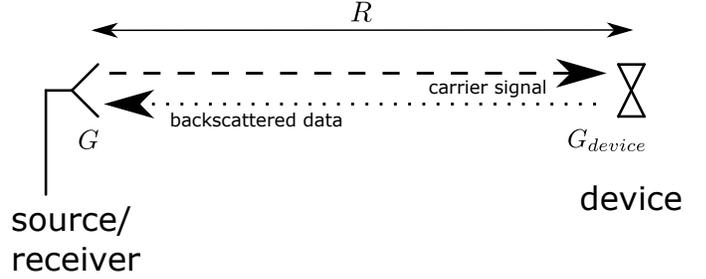}

\caption{Channel model of the mono-static setup: The source/receiver transmits
a carrier signal to the backscatter device (dashed arrow), which then
backscatters it back to the source/receiver (dotted arrow).}

\label{fig:channel_model_monostatic}
\end{figure}
Fig. \ref{eq:received_power_monostatic} shows the mono-static configuration,
where the source and the receiver use the same antenna. This simplifies
the calculations compared to the bi-static case, as $R=R_{C}=R_{RX}$.
Further, we can assume identical antenna gains for the source and
the reception antenna, i.e. $G=G_{C}=G_{RX}$.

\subsection{Path Loss Model}

As aforementioned, we focus on theoretical limits. Therefore, we neglect
channel conditions such as fading, shadowing, diffraction, and interference.
Instead, we describe the effects of the channel only by means of the
free-space path loss. Then, the power at the antenna output of the
passive device is given by \cite{Proakis2007}:

\begin{equation}
P_{RX,device}=G_{C}G_{device}\left(\frac{c_{0}}{4\pi R_{C}f_{c}}\right)^{2}P_{C},\label{eq:received_power_device}
\end{equation}
where $P_{C}$ is the transmit power of the carrier signal transmitted
by the source, $c_{0}$ is the speed of light, and $f_{c}$ is the
carrier frequency. We assume a system where the backscatter device
reflects the receiver power with efficiency $\mu$. This means the
transmit power of the backscatter device $P_{TX,device}$ is given
by $P_{TX,device}=\mu P_{RX,device}.$ Then, the power at the output
of the receiver antenna is given by:

\begin{equation}
P_{RX}=G_{RX}G_{device}\left(\frac{c_{0}}{4\pi R_{B}f_{c}}\right)^{2}\mu P_{RX,device}.\label{eq:received_power_BS_Friis}
\end{equation}
Using (\ref{eq:received_power_device}) in (\ref{eq:received_power_BS_Friis})
yields the received signal power $P_{RX,bs}$ at the receiver in a
bi-static setup, which is:

\begin{equation}
P_{RX,bs}=G_{RX}G_{C}G_{device}^{2}\left(\frac{c_{0}}{4\pi f_{c}}\right)^{4}\frac{\mu P_{C}}{R_{RX}^{2}R_{C}^{2}}.\label{eq:received_power_bistatic}
\end{equation}
In case of a mono-static setup, we use $R=R_{RX}=R_{C}$ and $G=G_{C}=G_{RX}$.
Then, (\ref{eq:received_power_bistatic}) simplifies to:

\begin{equation}
P_{RX,ms}=G^{2}G_{device}^{2}\left(\frac{c_{0}}{4\pi Rf_{c}}\right)^{4}\mu P_{C},\label{eq:received_power_monostatic}
\end{equation}
where $P_{RX,ms}$ is the received signal power at the receiver in
the mono-static case.

\subsection{Noise}

In order to model the noise we assume thermal noise with an ambient
temperature of $T=\unit[300]{K}$. This leads to a noise power spectral
density $N_{0,th}=\unit[-174]{dBm/Hz}$ \cite{molisch2012wireless}.
The noise is characterized by additive white Gaussian noise.

The total noise power $N$ is then given by $N=N_{0}W$, where $W$
is the bandwidth of the received signal and $N_{0}$ is the one-sided
noise power spectral density. $N_{0}$ is given by $N_{0}=N_{0,th}F$,
where $F$ is the receiver's noise figure. Therefore, the received
signal-to-noise ratio (SNR) is given by:

\begin{equation}
\mathrm{SNR}=\frac{P_{RX}}{N_{0,th}FW},\label{eq:SNR}
\end{equation}
where $P_{RX}$ is defined by (\ref{eq:received_power_bistatic})
for the bi-static configuration, and by (\ref{eq:received_power_monostatic})
for the mono-static configuration.

Assuming a correlation receiver, we can express the SNR also by the
$E_{b}/N_{0}$ \cite{Proakis2007}. The ratio between the received
energy per bit $E_{b}$, and the one-sided noise power spectral density
at the receiver $N_{0}$, in our system is obtained by considering
that $E_{b}=P_{RX}/r_{b}$, where $r_{b}$ is the data rate, and $N_{0}=N_{0,th}F$,
which yields: 
\begin{equation}
\frac{E_{b}}{N_{0}}=\frac{P_{RX}}{r_{b}N_{0,th}F}.\label{eq:Eb_N0}
\end{equation}

We are focusing on the theoretical bounds. Consequently, we will assume
a noise figure of $F=1$ (i.e. $\unit[0]{dB}$) in Sec. \ref{sec:Influence-of-Various}.
This means we assume a perfect receiver that does not add any additional
noise \cite{molisch2012wireless}. 

\subsection{Carrier Frequency, Bandwidth and Transmission Power\label{subsec:Carrier-Frequency,-Bandwidth}}

Due to the frequency regulation, the system parameters cannot be randomly
selected, especially in case of non-ambient backscattering. Available
bands for transmitting the carrier signal are the license-exempt ISM
(Industrial, Scientific and Medical) or SRD (Short Range Devices)
bands. As the backscatter signal typically has to remain in these
bands, its maximum bandwidth is limited.

For the US, the FCC (Federal Communications Commission) assigns three
ISM (Industrial, Scientific and Medical) bands around $\unit[915]{MHz}$
(with $\unit[26]{MHz}$ bandwidth), $\unit[2.4]{GHz}$ (with $\unit[83.5]{MHz}$
bandwidth), and $\unit[5.8]{GHz}$ (with $\unit[125]{MHz}$ bandwidth)
in FCC Title 47, §15.247 \cite{FCC15247}. The radiated electrical
power in these bands is limited to $\unit[1]{W}$ with an antenna
gain of up to $\unit[6]{dBi}$, leading to a maximum EIRP (effective
isotropic radiated power) of $\unit[4]{W}$ ($\unit[36]{dBm})$.

For Europe, the ETSI (European Telecommunications Standards Institute)
assigns two frequency bands for backscatter communications:
\begin{itemize}
\item ETSI EN 302 208 \cite{etsi_300220_2} defines a \textquotedbl lower
band\textquotedbl{} at $\unit[868]{MHz}$ with an ERP\footnote{The ERP is the antenna gain wrt. a half-wave dipole antenna. $\unit[1]{W}$
ERP is equivalent to $\unit[1.64]{W}$ EIRP.} (effective radiated power) of $\unit[2]{W}$ in a bandwidth of $\unit[200]{kHz}$,
and an \textquotedbl upper band\textquotedbl{} at $\unit[915]{MHz}$
with an ERP of $\unit[2]{W}$ in a bandwidth of $\unit[400]{kHz}$.
The latter, however, is only available in few European countries.
The ERP of $\text{\ensuremath{\unit[2]{W}}}$ is equivalent to an
EIRP of $\unit[3.28]{W}$.
\item ETSI EN 300 440 \cite{etsi_300440} assigns a $\unit[8]{MHz}$ wide
band at $\unit[2.4]{GHz}$ with an EIRP of $\text{\ensuremath{\unit[500]{mW}}}$,
or an EIRP of $\unit[4]{W}$ (with very strict restrictions to fixed
indoor setups).
\end{itemize}
Backscatter communication could also be used in licensed frequency
bands. An example is the use of FM radio or cellular networks signals
as ambient carrier signals. However, there currently exist no specific
frequency bands, channel bandwidths, or power limitations. Therefore,
we will mainly focus on the parameters available for license-exempt
frequency bands. 

\section{Derivation of the Theoretical Bound\label{sec:Derivation-of-the-Theoretical_Bound}}

Within this section we will derive the theoretical bounds for different
system assumptions.

\subsection{Absolute Bound}

According to Shannon's coding theory, error-free communication in
the AWGN (additive white Gaussian noise) channel is only possible
if ${E_{b}/N_{0}}_{min}\geq\ln(2)$ (or $\geq\unit[-1.59]{dB}$) \cite{Proakis2007}.
It has to be noted that this bound practically assumes an infinite
channel bandwidth $W$. Considering this in (\ref{eq:Eb_N0}), and
solving for $r_{b}$ obtains the theoretical bound $C_{\infty}$ for
the data rate in the infinite bandwidth channel with $r_{b}\leq C_{\infty}$: 

\begin{equation}
C_{\infty}=\frac{P_{RX}}{N_{0,th}F\ln(2)},\label{eq:absolute_bound}
\end{equation}
where $F$ is the noise figure, and $P_{RX}$ is given by (\ref{eq:received_power_bistatic})
for the bi-static, and by (\ref{eq:received_power_monostatic}) for
the mono-static configuration. We will call $C_{\infty}$ the ``absolute
bound'' as it assumes an infinite bandwidth.

\subsection{Bandwidth-limited Bound}

However, (\ref{eq:absolute_bound}) is only valid for systems with
an infinite channel bandwidth $W$, which may not be possible due
to the frequency regulation or practical aspects. For systems with
limited bandwidth, we have to use the capacity $C_{W}$ in the band-limited
AWGN channel with input power constraint, which is given by \cite{Proakis2007}:

\begin{equation}
C_{W}=W\log_{2}\left(1+\mathrm{SNR}\right).\label{eq:bound_for_limited_bandwidth}
\end{equation}
This bound depends on the bandwidth and on the $\mathrm{SNR}$, which
in our case is given by (\ref{eq:SNR}). Therefore, using (\ref{eq:SNR})
in (\ref{eq:bound_for_limited_bandwidth}) yields the bound for the
achievable data rate $r_{b}\leq C_{W}$ in the band-limited channel.
It is given by:

\begin{equation}
C_{W}=W\log_{2}\left(1+\frac{P_{RX}}{N_{0,th}FW}\right),\label{eq:data_rate_bound_limited_bandwidth}
\end{equation}
where $P_{RX}$ is again given by (\ref{eq:received_power_bistatic})
for the bi-static, and by (\ref{eq:received_power_monostatic}) for
the mono-static configuration. We will call $C_{W}$ the bandwidth-limited
bound.

\subsection{Effects of Realistic Channel Coding}

If realistic channel coding is applied to the transmission, there
will be additional losses compared to the bounds $C_{\infty}$ and
$C_{W}$. The capacity equations used in the previous sections assume
an infinite payload data length \cite{Dolinar1998}. However, the
typical information block size will be much smaller for typical backscatter
scenarios. Therefore, the limited code word length introduces an additional
loss given by the so-called Sphere Packing Bound. This topic is well-discussed
in \cite{Dolinar1998}, where \cite[Fig. 2]{Dolinar1998} shows the
resulting losses compared to the capacity for a word error-rate of
$10^{-4}$. The loss wrt. the capacity is more than $\unit[8]{dB}$
if a single payload bit is transmitted. For an information block size
of $\unit[100]{bits}$ we still have a loss of more then $\unit[2]{dB}$.
Furthermore, also the losses due to imperfect forward error correction
codes have to be taken into account. However, as we are focusing on
the maximum theoretical capacity, we will not consider the loss due
to realistic channel coding within the remaining part of this document.

\section{Achievable Data Rates for Different System Configurations\label{sec:Influence-of-Various}}

In this section we will examine the influence of different system
parameters on the theoretical capacities. This means we assume an
infinite information block size. Also, we assume a noise figure $F=\unit[0]{dB}$,
which characterizes the optimum receiver. For the passive device we
assume a backscatter efficiency $\mu=1$, which characterizes a perfect
backscatter device.

As the orientation of the passive backscatter device is unknown in
most cases, it is not useful to equip them with directive antennas.
Consequently, we assume a half-wave dipole with an antenna gain of
$G_{device}=\unit[2.15]{dBi}$. For the source/receiver antenna we
assume an antenna gain of $\unit[8]{dBi}$, which is a typical value
for UHF-RFID antennas. As the frequency regulation limits the maximum
EIRP to $\unit[4]{W}$ ($\unit[36]{dBm}$), we set the transmit power
to $P_{C}=\unit[0.63]{W}$ ($\unit[28]{dBm)}$. Furthermore, we assume
a carrier frequency of $f_{c}=\unit[900]{MHz}$. Finally, we assume
a mono-static backscatter configuration if not stated otherwise.

\subsection{Influence of the SNR\label{subsec:Influence-of-the}}

\begin{figure}[t]
\includegraphics[width=3.5in]{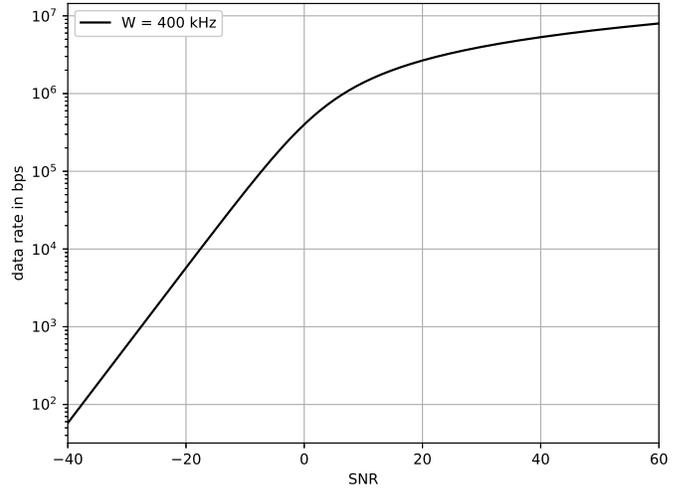}

\caption{Maximum data rate $r_{b}$ as a function of the $\mathrm{SNR}$ for
the absolute bound and the bandwidth-limited bound with $W=\unit[400]{kHz}$.}

\label{fig:SNR_vs_data_rate}%
\end{figure}
As a first step, we show the influence of the $\mathrm{SNR}$ on the
absolute bound the bandwidth-limited bound $C_{W}$ for a given bandwidth
$\unit[W=400]{kHz}$.

Fig. \ref{fig:SNR_vs_data_rate} shows a steep increase for the maximum
data rate $r_{b}$ if $\mathrm{SNR<\unit[0]{dB}}$, i.e. $\mathrm{SNR<1}$
in the linear case. This behavior arises from the log-term in (\ref{eq:bound_for_limited_bandwidth})
and (\ref{eq:data_rate_bound_limited_bandwidth}). This term can be
approximated by $\log_{2}(1+x)\thicksim x$, which means (\ref{eq:absolute_bound})
and (\ref{eq:data_rate_bound_limited_bandwidth}) are approximately
identical. This results in a linear increase with the $\mathrm{SNR}$.

For further increasing SNR the 1 in the log-term becomes negligible,
and the bandwidth limitation comes into effect. In this case we get
$\log_{2}(1+x)\thicksim10\log_{10}x$, where $10\log_{10}x$ is the
dB-value of $x$. Consequently, the capacity $C_{W}$ scales linearly
with the dB-value of the SNR. The actual transition point depends
on the ratio of $P_{RX}/(N_{0,th}FW)$ in (\ref{eq:data_rate_bound_limited_bandwidth}).
A higher bandwidth $W$ shifts the transition point to higher data
rates.

\subsection{Influence of the Signal Bandwidth }

\begin{figure}[t]
\includegraphics[width=3.5in]{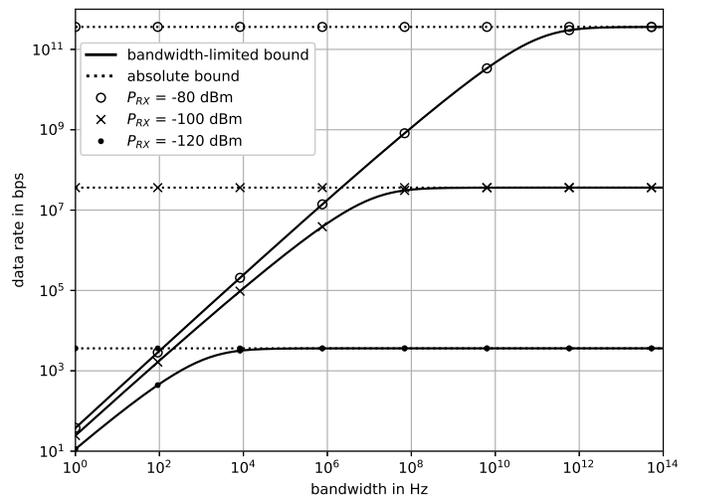}

\caption{Bound for the maximum data rate $r_{b}$ as a function of the bandwidth
$W$ for various signal powers $P_{RX}$, assuming $F=\unit[0]{dBm}$
and $N_{0,th}=\unit[-174]{dBm/Hz}$.}

\label{fig:data_rate_bandwidth}%
\end{figure}
Fig. \ref{fig:data_rate_bandwidth} shows the bandwidth-limited bound
for the maximum data rate $r_{b}$ in as a function of the bandwidth
$W$ for different signal reception powers $P_{RX}$. As dotted lines
show the absolute bounds assuming infinite bandwidth. We can see that
the bandwidth-limited bound for the data rate increases with increasing
bandwidth until a certain point. After that point there is no further
increase and the bandwidth-limited bound approximates to the absolute
bound. This is the same effect that we already discussed in Section
\ref{subsec:Influence-of-the}. It is the point, where the $\mathrm{SNR}$
in (\ref{eq:bound_for_limited_bandwidth}) reaches $\unit[0]{dB}$.
The location of this point depends both on the bandwidth and the power
of the received signal. It is reached later for higher signal powers.
Left hand side of this point the bandwidth $W$ is the main limiting
factor, resulting from the log-term in (\ref{eq:bound_for_limited_bandwidth}).
On the right hand side of this point the bandwidth $W$ is so high
that we can again approximate $\log_{2}(1+x)\approx x$ as $x<1$.
Hence, the SNR, or equivalently the received signal power $P_{RX}$
in (\ref{eq:data_rate_bound_limited_bandwidth}), becomes the limiting
factor.

\subsection{Influence of the Distance}

\begin{figure}[t]
\includegraphics[width=3.5in]{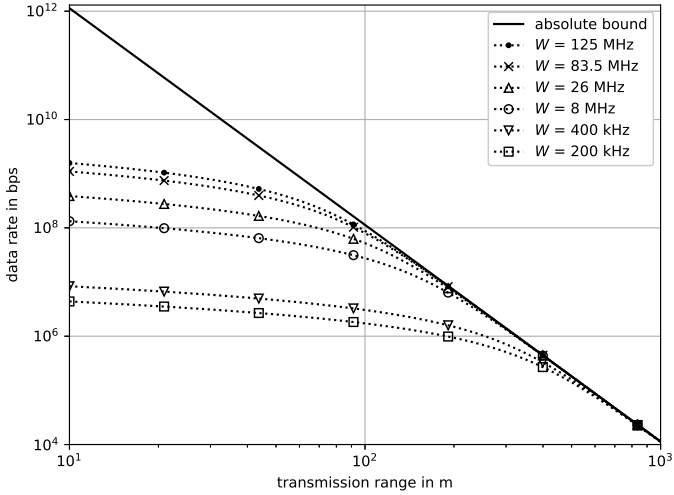}

\caption{Maximum data rate $r_{b}$ in a mono-static setup as a function of
the transmission range $R$. The solid curve shows the absolute bound
$C_{\infty}$, while non-solid curves show the bound $C_{W}$ for
the bandwidth-limited cases.}

\label{fig:data_rate_monostatic_bandwidth}%
\end{figure}
The effect of the distance $R$ is of high interest for many applications.
Fig. \ref{fig:data_rate_monostatic_bandwidth} shows the bandwidth-limited
$C_{W}$ bound for the maximum data rate $r_{b}$ for different bandwidths
as a function of the distance $R$. In addition, the solid curve shows
the absolute bound $C_{\infty}$ assuming an infinite bandwidth.

The figure shows that the data rate for bandwidth-limited systems
only decreases slowly as the distance increases. This is again in
the region where $\mathrm{SNR}>\unit[0]{dB}$, i.e. the log-term limits
the maximum data rate. For higher distances we get into the region
$\mathrm{SNR}<\unit[0]{dB}$, i.e. where we can approximate the log-term
by $\log_{2}(1+x)\approx x$. In this region, the system is not limited
by the bandwidth anymore, but by the low received signal power $P_{RX}$.
This also explains the full overlap with the absolute bound $C_{\infty}$
. In this region doubling the distance decreases the data rate by
the factor 16. The reason is given in (\ref{eq:received_power_monostatic}),
as $P_{RX}\sim R^{-4}$.

\subsection{Influence of Carrier Signal Power\label{subsec:Influence-of-Carrier-Signal_and_Noise_figure}}

An important system parameter in real systems is the carrier signal
power $P_{C}$, which is typically given by the frequency regulation.
Eq. (\ref{eq:SNR}) shows that it influences the received $\mathrm{SNR}$
linearly. Therefore, every increase of the carrier signal power by
$\text{\ensuremath{\unit[1]{dB}}}$ increases also the $P_{RX}$ \textendash{}
and equivalently the $\mathrm{SNR}$ \textendash{} by $\unit[1]{dB}$.
\begin{figure}[t]
\includegraphics[width=3.5in]{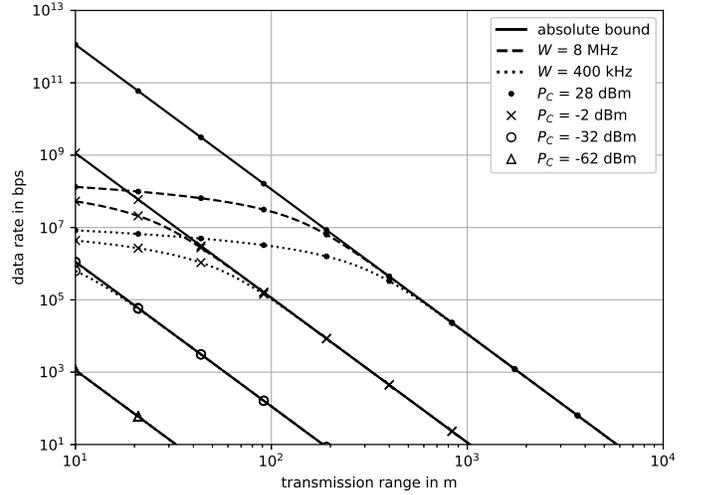}

\caption{Maximum data rate $r_{b}$ in a mono-static setup as a function of
the transmission range for different carrier signals powers $P_{C}$.
The solid curves show the absolute bound $C_{\infty}$, while the
dotted curves show the bandwidth-limited bound $C_{W}$.}

\label{fig:data_rate_monostatic_losses}%
\end{figure}

Fig. \ref{fig:data_rate_monostatic_losses} shows the influence of
$P_{C}$ on the maximum data rate $r_{b}$. We again assume the mono-static
configuration. Again we can observe the bandwidth limited regions
for high $\mathrm{SNR}$, indicated by the non-solid lines. For the
$\mathrm{SNR}$ limited region increasing the $P_{C}$ by $\unit[30]{dB}$
increases the data rate by the factor $1000$. The effect is much
smaller for the bandwidth limited region. Further, this figure sows
that the bandwidth $W$ causes a practical upper limit for the achievable
data rate, which is at around $\unit[150]{Mbps}$ for $W=\unit[8]{MHz}$,
and $\unit[10]{Mbps}$ for $W=\unit[400]{kHz}$. 

\subsection{Influence of the Carrier Frequency}

\begin{figure}[t]
\includegraphics[width=3.5in]{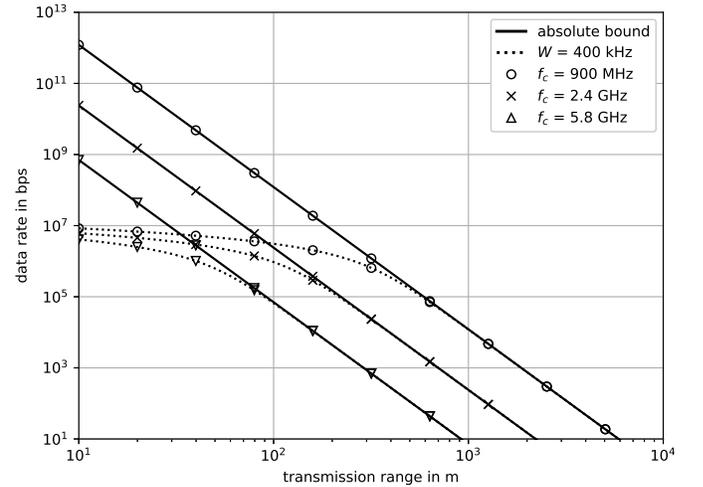}

\caption{Maximum bound for the data rate $r_{b}$ in a mono-static setup as
a function of the transmission range $R$ for different carrier frequencies
$f_{c}$. The solid curves show the absolute bound $C_{\infty}$,
while the dotted curves show the bandwidth-limited bound $C_{W}$
with $W=\unit[400]{kHz}$.}

\label{fig:data_rate_monostatic_fc}%
\end{figure}
Now we will analyze the effect of the carrier frequency $f_{c}$.
Fig. \ref{fig:data_rate_monostatic_fc} shows the absolute bound for
the data rate $r_{b}$ for the ISM band frequencies $f_{c}=\unit[915]{MHz}$,
$f_{c}=\unit[2.4]{GHz}$, and $f_{c}=\unit[5.8]{GHz}$. The dotted
line further shows the maximum data rate in case of a bandwidth of
$W=\unit[400]{kHz}$. We again use the mono-static configuration.
In the bandwidth limited case the maximum data rate again converges
to approx. $\unit[10]{Mbps}$. However, a strong impact of the carrier
frequency on the range $R$ is visible, as $P_{RX}\sim f_{c}^{-4}$.
Consequently, smaller frequencies are more suitable for backscatter
modulation, as they support higher ranges. The reason for this effect
is hidden in the free-space path loss equation. As we assume an omni-directional
pattern of the antenna of the passive device, we are limited to small
antenna gains $G_{device}$. Consequently, increasing the carrier
frequency will reduce the antenna size. Hence, this will also reduce
the effective antenna aperture (also called radar cross section) that
reflects the carrier signal to the receiver \cite{molisch2012wireless}.

\subsection{Influence of the Antenna Gain\label{subsec:Influence-of-the-antenna-gain}}

\begin{figure}[t]
\includegraphics[width=3.5in]{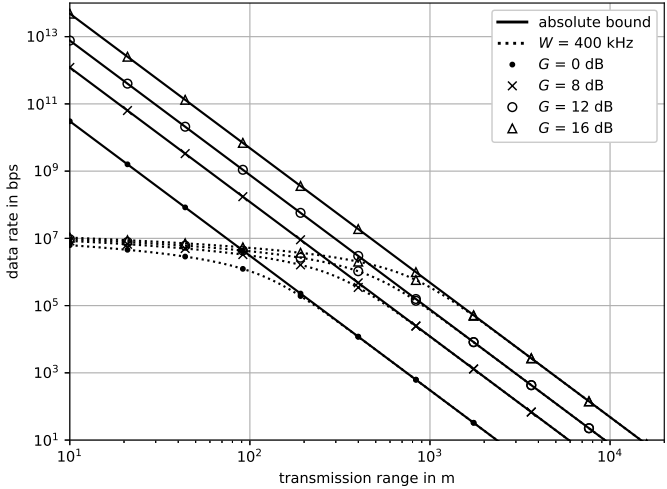}

\caption{Maximum data rate $r_{b}$ as a function of the transmission range
$R$ for various antenna gains $G$. The solid curves show the absolute
bound $C_{\infty}$, while the dotted curves show the bandwidth-limited
bound $C_{W}$ with $W=\unit[400]{kHz}$.}

\label{fig:data_rate_monostatic_antenna_gain}%
\end{figure}
Increasing the antenna gains is a classical method to improve the
received signal level, and hence the data rate or the distance. Here
we again assume the mono-static configuration. Consequently, the antenna
gain at the source and the receiver are identical, i.e. $G=G_{RX}=G_{C}$.
The antenna gain of the passive device is not modified.

Fig. \ref{fig:data_rate_monostatic_antenna_gain} shows the achievable
maximum data rates for different antenna gains $G$ as function of
the distance $R$. Again we assume a transmit carrier power of $P_{C}=\unit[28]{dBm}.$
Increasing the antenna gain helps on the transmit and receive side.
An increase of $\unit[1]{dB}$ improves the rate by the factor $1.58$,
or increases the range by factor $1.12$.

However, for practical applications we have to take the frequency
regulation into account that limits the EIRP $P_{C,EIRP}=P_{C}G$
(cf. Sec. \ref{subsec:Carrier-Frequency,-Bandwidth}). Consequently,
$P_{C}$ has to be reduced to meet the maximum allowed EIRP of e.g.
$P_{C,EIRP}=\unit[36]{dBm}$. This effectively means that an additional
antenna gain will only help on the receiver side. An additional antenna
gain of $\unit[1]{dB}$ will then improve the rate only by factor
$\sqrt{1.58}=1.26$, and the range by factor $\sqrt{1.12}=1.06$.

\subsection{Influence of the Distance in an Ambient Backscattering Configuration}

\begin{figure}[t]
\includegraphics[width=3.5in]{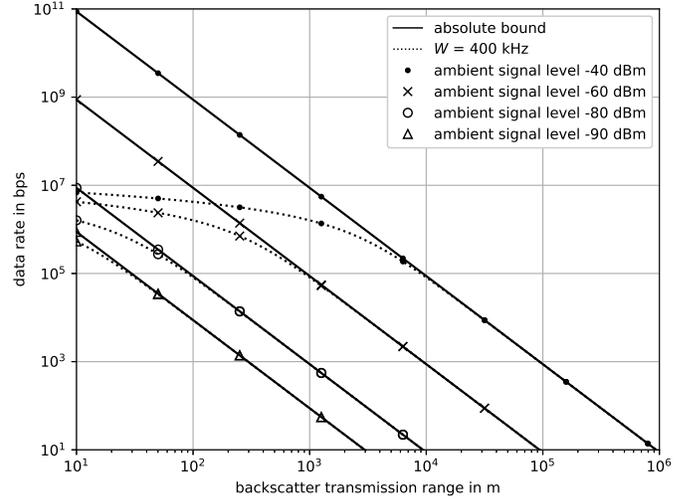}

\caption{Maximum data rate $r_{b}$ for a bi-static setup as a function of
the backscatter transmission range $R_{RX}$ for different ambient
signal levels at a carrier frequency of $f_{c}=\unit[200]{MHz}$.
The solid curves show the absolute bound $C_{\infty}$, while the
dotted curves show the bandwidth-limited bound $C_{W}$ with $W=\unit[400]{kHz}$.}

\label{fig:data_rate_bistatic}
\end{figure}

Here, we show the theoretical limits for ambient backscattering considering
Digital Audio Broadcasting (DAB) signals as ambient carrier signal.

DAB is a radio broadcast standard for digital radio service which
radiates a constant data stream without pauses. DAB systems are planned
in a way that enables a sufficient signal level for receiving DAB
radio service over a wide area in regions where it is deployed. The
guidelines for planning DAB networks of the European Broadcasting
Union (EBU) \cite{EBU} shows that the ambient signal level will almost
always be above $\unit[-90]{dBm}$. According to \cite{EBU}, planning
a DAB network with a carrier frequency of $\unit[200]{MHz}$ in such
a way that DAB service is available in 90 \% of the area in a rural
area requires a minimum median equivalent field strength of $\unit[38.64]{dB\lyxmathsym{\textmu}V/m}$
\cite[Table 24]{EBU}. Assuming that the backscatter device has an
antenna gain of $\unit[2.15]{dBi}$, this value relates to a received
ambient signal power at the backscatter device of $\unit[-82.6]{dBm}$.

Fig. \ref{fig:data_rate_bistatic} shows the theoretical limits for
the data rate for backscattering data over a certain transmission
distance $R_{RX}$ for various ambient signal levels. Compared to
dedicated source backscattering in monostatic setups, ambient backscattering
can - depending on the ambient signal level - achieve much higher
data rates. This is particularly the case for large transmission distances,
where the ambient signal level will usually be much higher than the
dedicated carrier that has to be transmitted over the same distance.
Another advantage is, that there are ambient signals at lower frequencies
than the frequencies we identified for dedicated source backscattering,
which leads to a lower path loss.

\subsection{Influence of the Carrier Signal Uptime}

Especially in case of ambient scenarios the carrier signal uptime
may be limited. Examples are WiFi beacons or reference signals in
cellular networks. Typical WiFi access points transmit beacon signals
approx. every $\unit[100]{ms}$ that have a duration in the order
of one to few milliseconds \cite{perahia2013next}. The 4G cellular
standard LTE transmits the Primary and Secondary Reference Signals
every $\unit[5]{ms}$ that have a duration in the order of $\unit[130]{\text{\textmu s}}$
\cite{dahlman20185g}. In 5G, the Synchronization Signal Block is
typically transmitted every $\unit[20]{ms}$ and has a duration of
$\thickapprox\unit[285]{\lyxmathsym{\textmu}s}$ (for frequency bands
$<\unit[3]{GHz}$) \cite{dahlman20185g}.

\begin{figure}[t]
\includegraphics[width=3.5in]{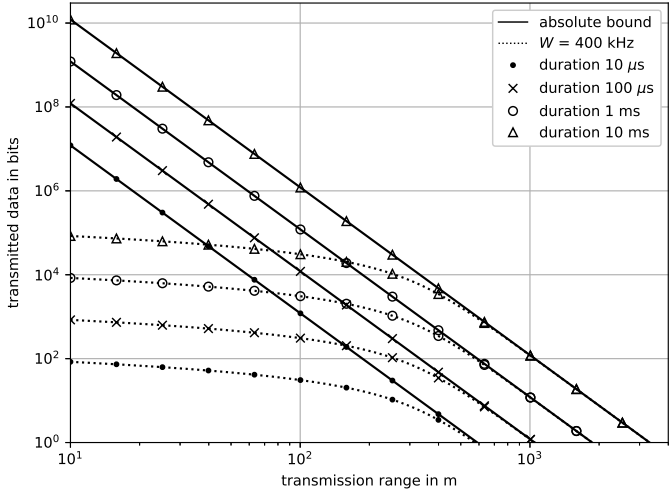}

\caption{Maximum number of data bits that can be transmitted in the given carrier
signal uptime as a function of the transmission range $R$.}

\label{fig:required_data_rate}%
\end{figure}
Fig. \ref{fig:required_data_rate} shows the maximum amount of data
that can be transmitted within a given signal uptime as function of
the transmission range $R$. Here, we again assume the mono-static
configuration. Already a signal uptime in the order of $\unit[1]{ms}$
offers the possibility to transmit sufficient number of payload bits.
This also holds for the bandwidth limited case with $W=\unit[400]{kHz}$.

\section{Application of the Bound on UHF-RFID Systems\label{sec:Application-of-the-RIFD}}

We can now use the derived bound to compare them against state-of-the-art
UHF-RFID \cite{Spec} systems. For this purpose we assume a mono-static
RFID reader operating according to FCC 15.247 rules \cite{FCC15247}:
The reader transmits a continuous wave with a transmit power of $P_{C}=\unit[28]{dBm}$.
Further, it uses an antenna gain of $G=\unit[8]{dBi}$ and a carrier
frequency of $f_{c}=\unit[915]{MHz}$. This results in the maximum
allowed EIRP of $\unit[36]{dBm}$ (i.e. $\unit[4]{W}$). For the backscatter
device we use assume a backscatter efficiency of $\mu=25\%$ \cite{Kuester2013HowGood},
and an antenna gain of $G_{device}=\unit[2.15]{dBi}$. Furthermore,
we assume the two distances of $R=\unit[10]{m}$ and $R=\unit[100]{m}$.
In addition, we use $F=\unit[20]{dB}$ to represent the losses due
to the receiver noise figure (typically high due to full duplex operation),
and the transmission of uncoded data bits as UHF RFID does not use
a forward error correction code.

We can now use (\ref{eq:received_power_monostatic}) to calculate
the received power, and (\ref{eq:absolute_bound}) to calculate the
maximum rate for the absolute bound. We do not have to calculate the
bound for the bandwidth limited case, as this bound will be very close
to the absolute bound. This is caused by the high bandwidth of $\unit[26]{MHz}$
in the regarded frequency range.

\begin{table}[H]
\caption{\label{tab:RFID-example}Bound applied to UHF-RFID}

\centering{}%
\begin{tabular}{|c||c|c|}
\hline 
 & $R=\unit[10]{m}$ & $R=\unit[100]{m}$\tabularnewline
\hline 
\hline 
$P_{RX}$ & $\unit[-63]{dBm}$ & $\unit[-103]{dBm}$\tabularnewline
\hline 
$C_{\infty}$ & $\unit[18.2]{Mbps}$ & $\unit[18.2]{kbps}$\tabularnewline
\hline 
\end{tabular}
\end{table}
Table \ref{tab:RFID-example} shows the resulting values. These are
quite close to the values supported by UHF RFID. Using battery assisted
RFID transponders a reading range of up to $\unit[100]{m}$ is supported\footnote{https://www.impinj.com/products/technology/how-can-rfid-systems-be-categorized
(accessed Dec. 02, 2022)}. The minimum data rate offered by UHF RFID is $\unit[5]{kbps}$ if
Miller-8 encoding is used \cite{Spec}, and hence smaller than the
bound of $C_{\infty}=\unit[18.2]{kbps}$ for $R=\unit[100]{m}$. In
contrast, the highest rate offered by UHF RFID is $\unit[640]{kbps}$,
which is smaller than the bound of $C_{\infty}=\unit[18.2]{Mbps}$
for $R=\unit[10]{m}$. In summary, these calculated values fit quite
well to real UHF RFID backscatter systems.

\section{Summary and Conclusions\label{sec:Conclusion}}

Within this paper we developed theoretical capacity bounds for backscatter
communications. After defining the backscatter channel model we developed
the theoretical bounds based on classical information theory. We especially
analyzed the impact of the transmission power, the antenna gains,
the carrier frequency, and the system configuration on the achievable
maximum data rates in backscatter scenarios. The results show that
backscatter communication is able to support high data rates at relevant
transmit ranges. We have also shown that the maximum signal bandwidth
is an important factor in case of higher signal levels.

Backscatter communication is able to achieve payload data rates in
the order of several Mbps. Additionally, also long-range communication
over hundreds of meters is possible, but with significantly lower
payload data rates. One main limiting factor is the frequency regulation
that limits the maximum transmit power for the carrier signals if
license-exempt frequency bands are used. Finally, we have shown that
our theoretical bounds predict the performance of state-of-the-art
UHF RFID systems.

Within the scope of this work we have not made any assumption on the
modulation, with the exception of the total bandwidth. Therefore,
our future work will focus on efficient backscatter waveform designs
that are able to exploit the theoretical capacities as close as possible.

\section*{Acknowledgment}

This work is part of the research project 5G-Flexi-Cell (grant no.
01MC22004B) funded by the German Federal Ministry for Economic Affairs
and Climate Action (BMWK) based on a decision taken by the German
Bundestag.

\bibliographystyle{IEEEtran}

\begin{thebibliography}{10}
\providecommand{\url}[1]{#1}
\csname url@samestyle\endcsname
\providecommand{\newblock}{\relax}
\providecommand{\bibinfo}[2]{#2}
\providecommand{\BIBentrySTDinterwordspacing}{\spaceskip=0pt\relax}
\providecommand{\BIBentryALTinterwordstretchfactor}{4}
\providecommand{\BIBentryALTinterwordspacing}{\spaceskip=\fontdimen2\font plus
\BIBentryALTinterwordstretchfactor\fontdimen3\font minus
  \fontdimen4\font\relax}
\providecommand{\BIBforeignlanguage}[2]{{%
\expandafter\ifx\csname l@#1\endcsname\relax
\typeout{** WARNING: IEEEtran.bst: No hyphenation pattern has been}%
\typeout{** loaded for the language `#1'. Using the pattern for}%
\typeout{** the default language instead.}%
\else
\language=\csname l@#1\endcsname
\fi
#2}}
\providecommand{\BIBdecl}{\relax}
\BIBdecl

\bibitem{BackscatteringTorres2021}
R.~Torres, R.~Correia, N.~B. Carvalho, S.~Daskalakis, G.~Goussetis, Y.~Ding,
  A.~Georgiadis, A.~Eid, J.~Hester, and M.~M. Tentzeris, ``Backscatter
  communications,'' \emph{IEEE Journal of Microwaves}, vol.~1, no.~4, pp.
  864--878, 2021.

\bibitem{xu2018practical}
C.~Xu, L.~Yang, and P.~Zhang, ``Practical backscatter communication systems for
  battery-free internet of things: A tutorial and survey of recent research,''
  \emph{IEEE Signal Processing Magazine}, vol.~35, no.~5, pp. 16--27, 2018.

\bibitem{9145564}
I.~F. Akyildiz, A.~Kak, and S.~Nie, ``{6G} and beyond: The future of wireless
  communications systems,'' \emph{IEEE Access}, vol.~8, pp. 133\,995--134\,030,
  2020.

\bibitem{Nawaz2021}
S.~J. Nawaz, S.~K. Sharma, B.~Mansoor, M.~N. Patwary, and N.~M. Khan,
  ``Non-coherent and backscatter communications: Enabling ultra-massive
  connectivity in 6g wireless networks,'' \emph{IEEE Access}, vol.~9, 2021.

\bibitem{FinkenzellerEnglisch}
K.~Finkenzeller, \emph{RFID Handbook: Fundamentals and Applications in
  Contactless Smart Cards, Radio Frequency Identification and Near-Field
  Communication}, 3rd~ed.\hskip 1em plus 0.5em minus 0.4em\relax John Wiley \&
  Sons, Ltd, 2010.

\bibitem{BackscatterModulatorEnergieverbrauch}
R.~Correia and N.~B. Carvalho, ``Ultrafast backscatter modulator with low-power
  consumption and wireless power transmission capabilities,'' \emph{IEEE
  Microwave and Wireless Components Letters}, vol.~27, no.~12, pp. 1152--1154,
  2017.

\bibitem{Spec}
``{Class-1 Generation-2 UHF RFID Protocol for Communications at 860 MHz - 960
  MHz},'' EPC Global, 2016.

\bibitem{AmbientBackscatterSurveyHuynh}
N.~Van~Huynh, D.~T. Hoang, X.~Lu, D.~Niyato, P.~Wang, and D.~I. Kim, ``Ambient
  backscatter communications: A contemporary survey,'' \emph{IEEE
  Communications Surveys \& Tutorials}, vol.~20, no.~4, pp. 2889--2922, 2018.

\bibitem{8057162}
C.~Yang, J.~Gummeson, and A.~Sample, ``Riding the airways: Ultra-wideband
  ambient backscatter via commercial broadcast systems,'' in \emph{IEEE INFOCOM
  2017 - IEEE Conference on Computer Communications}, 2017, pp. 1--9.

\bibitem{Talla2017}
V.~Talla, M.~Hessar, B.~Kellogg, A.~Najafi, J.~R. Smith, S.~Gollakota, and
  P.~G. Allen, ``{LoRa} backscatter: Enabling the vision of ubiquitous
  connectivity,'' \emph{Proc. ACM Interact. Mob. Wearable Ubiquitous Technol},
  vol.~1, 2017.

\bibitem{Chen2021}
S.~Chen, M.~Zhang, J.~Zhao, W.~Gong, and J.~Liu, ``Reliable and practical
  {Bluetooth} backscatter with commodity devices,'' \emph{IEEE/ACM Transactions
  on Networking}, vol.~29, pp. 1717--1729, 8 2021.

\bibitem{201552}
A.~Wang, V.~Iyer, V.~Talla, J.~R. Smith, and S.~Gollakota, ``{FM} backscatter:
  Enabling connected cities and smart fabrics,'' in \emph{14th USENIX Symposium
  on Networked Systems Design and Implementation (NSDI 17)}.\hskip 1em plus
  0.5em minus 0.4em\relax Boston, MA: USENIX Association, mar 2017, pp.
  243--258.

\bibitem{8058860}
S.-N. Daskalakis, J.~Kimionis, A.~Collado, M.~M. Tentzeris, and A.~Georgiadis,
  ``Ambient fm backscattering for smart agricultural monitoring,'' in
  \emph{2017 IEEE MTT-S International Microwave Symposium (IMS)}, 2017, pp.
  1339--1341.

\bibitem{9295547}
R.~Torres, R.~Correia, and N.~B. Carvalho, ``All digital ambient backscatter
  system,'' in \emph{2020 IEEE Wireless Power Transfer Conference (WPTC)},
  2020, pp. 327--330.

\bibitem{9513239}
H.~Hwang, R.~B. Nti, and J.-H. Yun, ``Spectro-temporal combining in bistate
  {WiFi} backscatter communication with frequency shift,'' \emph{IEEE Access},
  vol.~9, pp. 113\,735--113\,747, 2021.

\bibitem{KelloggWIFIBackscattering}
B.~Kellogg, A.~Parks, S.~Gollakota, J.~R. Smith, and D.~Wetherall, ``{Wi-Fi}
  backscatter: Internet connectivity for rf-powered devices,'' in
  \emph{Proceedings of the 2014 ACM Conference on SIGCOMM}, ser. SIGCOMM
  2014.\hskip 1em plus 0.5em minus 0.4em\relax New York, NY, USA: Association
  for Computing Machinery, 2014, pp. 607--618.

\bibitem{9488716}
Q.~Wang, S.~Chen, J.~Zhao, and W.~Gong, ``Rapidrider: Efficient {WiFi}
  backscatter with uncontrolled ambient signals,'' in \emph{IEEE INFOCOM 2021 -
  IEEE Conference on Computer Communications}, 2021, pp. 1--10.

\bibitem{8647245}
M.~A. ElMossallamy, Z.~Han, M.~Pan, R.~J{\"a}ntti, K.~G. Seddik, and G.~Y. Li,
  ``Backscatter communications over ambient {OFDM} signals using null
  subcarriers,'' in \emph{2018 IEEE Global Communications Conference
  (GLOBECOM)}, 2018, pp. 1--6.

\bibitem{8103807}
G.~Yang, Y.-C. Liang, R.~Zhang, and Y.~Pei, ``Modulation in the air:
  Backscatter communication over ambient {OFDM} carrier,'' \emph{IEEE
  Transactions on Communications}, vol.~66, no.~3, pp. 1219--1233, 2018.

\bibitem{LinkBudgetPark2021}
B.~Park and H.-G. Ryu, ``Link budget investigation of ambient backscatter
  communication,'' in \emph{2021 IEEE International Conference on Consumer
  Electronics-Asia (ICCE-Asia)}, 2021, pp. 1--3.

\bibitem{8885700}
B.~Badihi, A.~Liljemark, M.~U. Sheikh, J.~Lietz\'{e}n, and R.~J{\"a}ntti,
  ``Link budget validation for backscatter-radio system in sub-{1GHz},'' in
  \emph{2019 IEEE Wireless Communications and Networking Conference (WCNC)},
  2019, pp. 1--6.

\bibitem{7948789}
R.~Duan, R.~J{\"a}ntti, H.~Yi\v{g}itler, and K.~Ruttik, ``On the achievable
  rate of bistatic modulated rescatter systems,'' \emph{IEEE Transactions on
  Vehicular Technology}, vol.~66, no.~10, pp. 9609--9613, 2017.

\bibitem{AmbientBackscatterZhao2018}
W.~Zhao, G.~Wang, R.~Fan, L.~S. Fan, and S.~Atapattu, ``Ambient backscatter
  communication systems: Capacity and outage performance analysis,'' \emph{IEEE
  Access}, vol.~6, 2018.

\bibitem{9051982}
Y.~Ye, L.~Shi, X.~Chu, and G.~Lu, ``On the outage performance of ambient
  backscatter communications,'' \emph{IEEE Internet of Things Journal}, vol.~7,
  no.~8, pp. 7265--7278, 2020.

\bibitem{9055221}
L.~Shi, R.~Q. Hu, Y.~Ye, and H.~Zhang, ``Modeling and performance analysis for
  ambient backscattering underlaying cellular networks,'' \emph{IEEE
  Transactions on Vehicular Technology}, vol.~69, no.~6, pp. 6563--6577, 2020.

\bibitem{Proakis2007}
Proakis, \emph{Digital Communications 5th Edition}.\hskip 1em plus 0.5em minus
  0.4em\relax McGraw Hill, 2007.

\bibitem{Kuester2013HowGood}
D.~Kuester and Z.~Popovic, ``How good is your tag?: {RFID} backscatter metrics
  and measurements,'' \emph{IEEE Microwave Magazine}, vol.~14, no.~5, pp.
  47--55, 2013.

\bibitem{molisch2012wireless}
A.~F. Molisch, \emph{Wireless communications}.\hskip 1em plus 0.5em minus
  0.4em\relax John Wiley \& Sons, 2012.

\bibitem{FCC15247}
\BIBentryALTinterwordspacing
FCC. {Title 47}, 15.247 {Operation} within the bands 902-928 {MHz}, 2400-2483.5
  {MHz}, and 5725-5850 {MHz}.
\BIBentrySTDinterwordspacing

\bibitem{etsi_300220_2}
ETSI, \emph{Short Range Devices (SRD) operating in the frequency range 25 MHz
  to 1 000 MHz; Part 2: Harmonised Standard for access to radio spectrum for
  non specific radio equipment}, EN 300 220-2, Rev. 3.2.1, Jun. 2018.

\bibitem{etsi_300440}
------, \emph{{Short Range Devices (SRD); Radio} equipment to be used in the {1
  GHz to 40 GHz} frequency range; {Harmonised} Standard covering the essential
  requirements of article 3.2 of {Directive 2014/53/EU}}, EN 300 440, Rev.
  2.2.1, Jul. 2018.

\bibitem{Dolinar1998}
S.~{Dolinar}, D.~{Divsalar}, and F.~{Pollara}, ``{Code Performance as a
  Function of Block Size},'' \emph{Telecommunications and Mission Operations
  Progress Report}, vol. 133, pp. 1--23, Jan. 1998.

\bibitem{EBU} 
\emph{Tech 3391 - Guidelines for DAB Network Planning}, European Broadcasting Union, Brussels, Belgium, May 2018. [Online]. Available: https://tech.ebu.ch/docs/tech/tech3391.pdf

\bibitem{perahia2013next}
E.~Perahia and R.~Stacey, \emph{Next generation wireless {LANs}: 802.11 n and
  802.11 ac}.\hskip 1em plus 0.5em minus 0.4em\relax Cambridge university
  press, 2013.

\bibitem{dahlman20185g}
E.~Dahlman, S.~Parkvall, and J.~Skold, \emph{5G NR: The Next Generation
  Wireless Access Technology}.\hskip 1em plus 0.5em minus 0.4em\relax Elsevier
  Science, 2018.
  

\end{thebibliography}

\begin{IEEEbiography}[{\includegraphics[width=1in,height=1.25in]{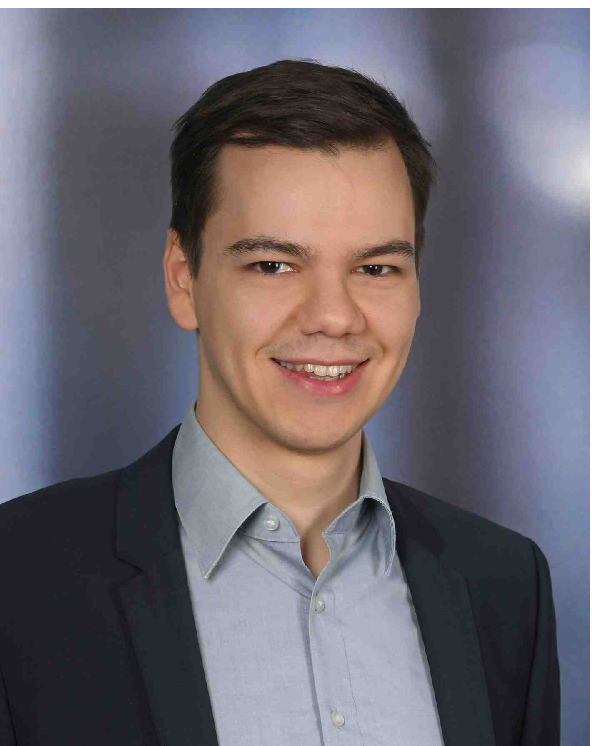}}]{Clemens Korn}
 received the B.Sc. and M.Sc. degree in Electrical Engineering from
Friedrich-Alexander University Erlangen-Nuremberg in 2015 and 2017,
respectively. In 2017 he joined Fraunhofer IIS, where he conducted
research and various industry projects with focus on UHF-RFID and
other IoT Systems. Further, he was involved in 3GPP standardization
for Fraunhofer IIS as RAN1 delegate for the RedCap study item. In
2021, he started a doctorate at TU Ilmenau, where he conducts research
on Zero-Energy Communications.
\end{IEEEbiography}

\begin{IEEEbiography}[{\includegraphics[width=1in,height=1.25in]{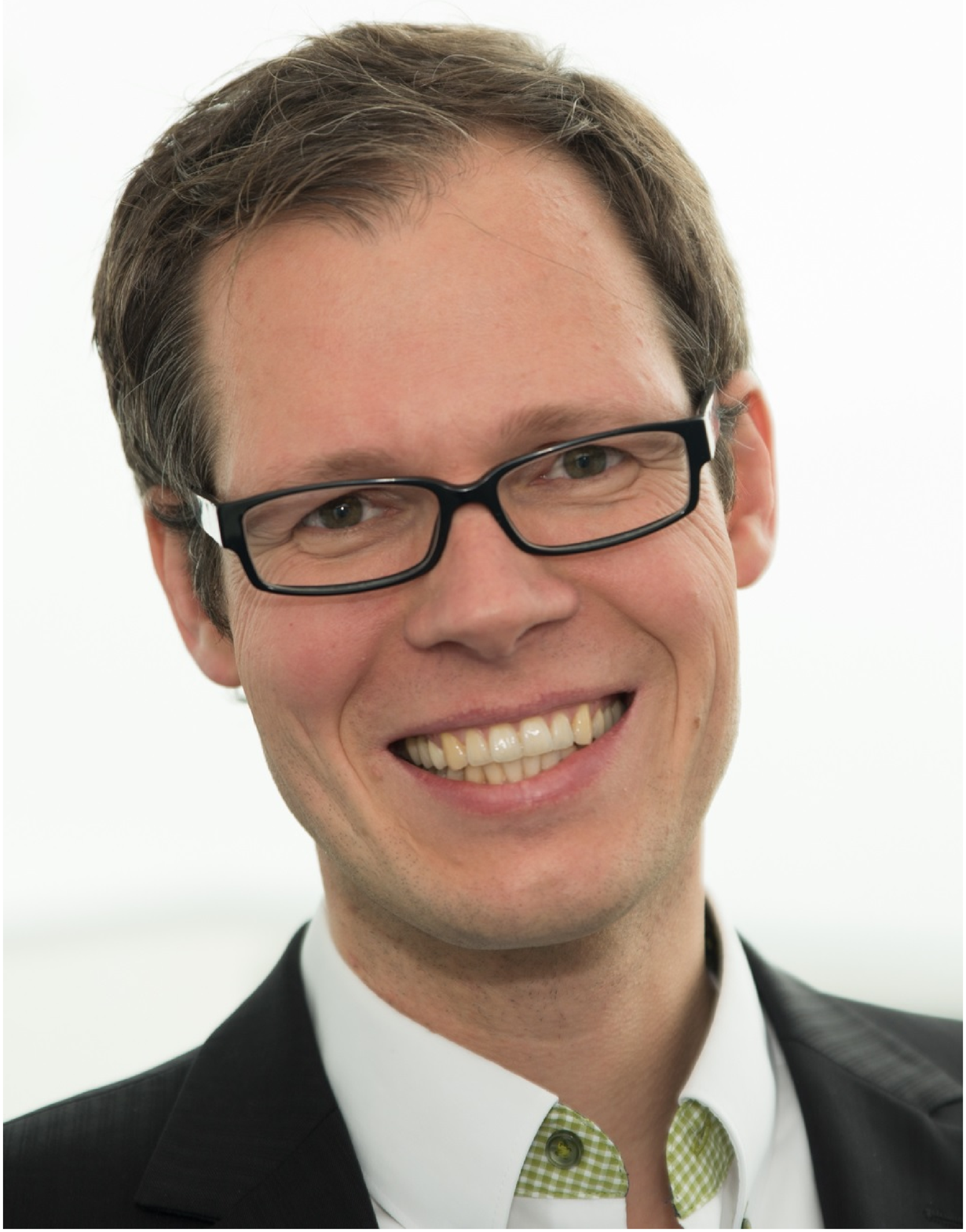}}]{Joerg Robert}
 studied electrical engineering and information technology at TU
Ilmenau and TU Braunschweig. From 2006 to 2012, he conducted research
on the topic of digital television at the Institute of Communications
Engineering at TU Braunschweig. Here he was actively involved in the
development of DVB-T2, the second generation of digital terrestrial
television. In 2013, he completed his doctorate at the TU Braunschweig
on the topic of \textquotedbl Terrestrial TV Broadcast using Multi-Antenna
Systems\textquotedbl . In 2012, he moved to the LIKE chair at the
Friedrich-Alexander-Universität Erlangen-Nürnberg. One of his research
focuses there was LPWAN (Low Power Wide Area Networks) and he led
the international standardization of LPWAN within IEEE 802. Since
2021, he is university professor and head of the Group for Dependable
Machine-to-Machine Communication at the Department of Electrical Engineering
and Information Technology at Technische Universität Ilmenau, Germany.
\end{IEEEbiography}

\begin{IEEEbiography}[{\includegraphics[width=1in,height=1.25in]{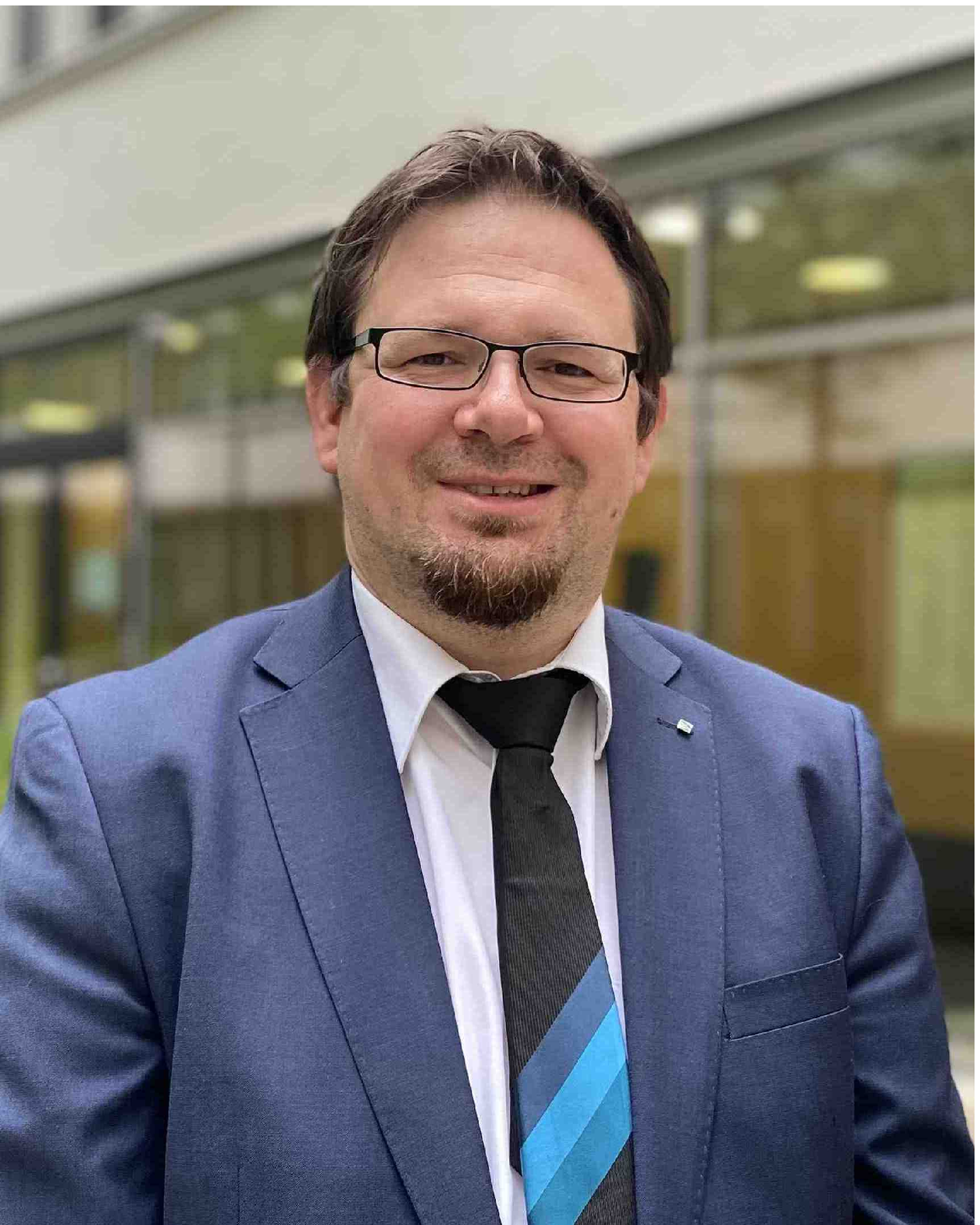}}]{Tobias Dräger}
 studied electrical engineering at the Friedrich-Alexander University
Erlangen-Nuremberg, specializing in automation technology/sensor technology.
After graduating in 2007, he worked at the Fraunhofer Institute for
Integrated Circuits IIS with a focus on RFID technology and its application,
embedded hardware and wireless transmission technologies. Since 2018,
he is head of the \textquotedbl RFID and inductive sensor systems\textquotedbl{}
group, which, in addition to the topics of embedded hardware, wireless
energy and data transmission, LTE-Cat NB1 and RFID, also works on
the topic of inductive near-field localization IndLoc®.
\end{IEEEbiography}

\enlargethispage{-0cm} 
\end{document}